\documentclass[twocolumn,aps,prl]{revtex4}

\usepackage{amssymb}
\usepackage{graphicx}
\usepackage{graphicx,amsfonts}
\usepackage{bm}

\begin{document}

\title{Observation of the spontaneous vortex phase in the weakly
ferromagnetic superconductor ErNi$_{2}$B$_{2}$C: A penetration depth study}
\author{Elbert E. M. Chia}
\affiliation{Department of Physics, University of Illinois at Urbana-Champaign, 1110 W.
Green Street, Urbana Illinois 61801}
\author{M. B. Salamon}
\affiliation{Department of Physics, University of Illinois at Urbana-Champaign, 1110 W.
Green Street, Urbana Illinois 61801}
\author{Tuson Park}
\affiliation{Los Alamos National Laboratory, MST-10, Los Alamos, New Mexico 87545}
\author{Heon-Jung Kim}
\author{Sung-Ik Lee}
\affiliation{National Creative Research Initiative Center for Superconductivity and
Department of Physics, Pohang University of Science and Technology, Pohang
790-784, Republic of Korea}
\author{Hiroyuki Takeya}
\affiliation{National Institute for Materials Science, 1-2-1 Sengen, Tsukuba, Ibaraki
305-0047, Japan}
\date{\today}

\begin{abstract}
The coexistence of weak ferromagnetism and superconductivity in ErNi$_{2}$B$%
_{2}$C suggests the possibility of a spontaneous vortex phase (SVP) in which
vortices appear in the absence of an external field. We report evidence for
the long-sought SVP from the in-plane magnetic penetration depth $\Delta
\lambda (T)$ of high-quality single crystals of ErNi$_{2}$B$_{2}$C. In
addition to expected features at the N\'{e}el temperature $T_{N}$ = 6.0~K
and weak ferromagnetic onset at $T_{WFM}=2.3~$K, $\Delta \lambda (T)$ rises
to a maximum at $T_{m}=0.45$~K before dropping sharply down to $\sim $0.1~K.
We assign the 0.45~K-maximum to the proliferation and freezing of
spontaneous vortices. A model proposed by Koshelev and Vinokur explains the
increasing $\Delta \lambda (T)$ as a consequence of increasing vortex
density, and its subsequent decrease below $T_{m}$ as defect pinning
suppresses vortex hopping.
\end{abstract}

\maketitle

It is now clear that the borocarbide superconductor ErNi$_{2}$B$_{2}$C
develops weak ferromagnetism (WFM) below T$_{WFM}$ = 2.3~K while remaining a
singlet superconductor \cite{Choi01,Canfield96}. The question naturally
arises: how do these two seemingly incompatible orders --- ferromagnetism
and superconductivity --- coexist microscopically? Clearly superconductivity
will be suppressed if the internal field $B_{in}$ generated by the
ferromagnetic moment exceeds $H_{c}$ for a Type-I, or $H_{c2}$ for a
Type-II, superconductor (SC). For a Type-II SC, however, vortices are
predicted to appear \textit{spontaneously} if $B_{in}$ lies in the range $%
H_{c1}<B_{in}\sim 4\pi M<H_{c2}$ \cite%
{Blount79,Tachiki79,Greenside81,Kuper80}. In this spontaneous vortex phase
(SVP), the vortex screening currents shield superconducting regions from the
intrinsic magnetization. The vortices, however, may be qualitatively
different from those generated by externally applied fields \cite%
{radzihovsky01}. In this Letter we report unusual features in the
penetration depth data of a high quality single-crystal of ErNi$_{2}$B$_{2}$%
C that give strong evidence for the existence of the SVP.

There have been previous SVP reports that we consider inconclusive. Ng and
Varma \cite{Ng97b}, for example, interpreted small angle neutron scattering
(SANS) data on ErNi$_{2}$B$_{2}$C as a prelude to the SVP. \ In that
experiment, Yaron et al. \cite{Yaron96} reported that the vortex-line
lattice begins to tilt away from the \textit{c}-axis (along which the
magnetic is field applied) towards the \textit{a-b} plane below $T_{WFM}$.
However, the tilt can merely be a result of the vector sum of the applied
field and the internal field produced by the ferromagnetic domains in the
basal plane. Additional evidence was provided by SANS data \cite{Furukawa01}
with the applied magnetic field in the basal plane. A large field was
applied to align ferromagnetic domains. When the field was removed, the flux
line lattice was found to persist below $T_{WFM}$ but disappear above it.
However, owing to the low $T_{c}$ ($\sim $ 8.5~K) and increased pinning
below $T_{WFM}$ \cite{Gammel00}, trapped flux cannot be ruled out.

Among the magnetic members of the rare-earth (RE) nickel borocarbide family,
RENi$_{2}$B$_{2}$C (RE = Ho, Er, Dy, etc.), ErNi$_{2}$B$_{2}$C, is a
particularly good candidate for study. Superconductivity arises at $%
T_{c}\approx ~11$~K and persists when antiferromagnetic (AF) order sets in
at $T_{N}$ $\approx $~6~K \cite{Cho95}. In the AF state the Er spins are
directed along the \textit{b}-axis, forming a transversely polarized,
incommensurate sinusoidal spin-density-wave (SDW) state, with modulation
vector modulation vector $\delta $ = 0.553$a^{\ast }$ (a$^{\ast }$ = 2$\pi $/%
$a$) \cite{Zarestky95}. The appearance of higher-order reflections at lower
temperatures \cite{Choi01} signals the development of a square-wave
modulation, with regular spin slips spaced by 20$a$. Below 2.3~K WFM appears
with $B_{in}\cong $0.1~T, approximately one Er magnetic moment per twenty
unit cells, clearly correlated with spin slips.

The relative stability of various phases of a ferromagnetic
superconductor was explored \cite{Greenside81} by Greenside
\textit{et al.} A spiral phase is not possible in the presence of
strong uniaxial anisotropy and the spontaneous vortex phase is more
stable than a linearly polarized state for small values of $\zeta
=[F_{FM}/F_{s}]$, the ratio of\ ferromagnetic to superconducting
free-energy densities at $T=0$. For $\zeta =100$ and the ratio
$\lambda /\gamma =10$, where $\gamma
=[3k_{B}T_{c}S/(2aM^{2}(S+1)]^{1/2}$ is related to the exchange
stiffness, Greenside \textit{et al.} find the SVP to be the most
stable low-temperature phase; indeed, they suggest that the effect
is most likely to be found in a dilute ferromagnetic superconductor,
and that smaller values of $\zeta $ favor SVP. In the case of
ErNi$_{2}$B$_{2}$C, where only 5\% of the Er atoms contribute to
ferromagnetism, we have $F_{s}=-H_{c}^{2}/8\pi \approx -1.5\times
10^{5}$
erg/cm$^{3}$, where $H_{c}$$\approx $ 1900~G from Ref.~%
\onlinecite{Canfield96}. The ferromagnetic energy density is $%
F_{FM}=-3Nk_{B}T_{c}S/(2(S+1))\approx -4.3\times 10^{6}$ erg/cm$^{3}$, where
$N$=1.5x10$^{22}$~cm$^{-3}$ is the density of the (magnetic) Er atoms, and $%
S=$ 3/2 is the Er spin. This then gives $\zeta =30$, strongly favoring the
SVP. The spin-stiffness length is $\gamma =100$~\AA\ at low temperatures
where $M\approx $ 88 G, so that $\lambda /\gamma \approx 7,$ close to the
value assumed in Ref.~\onlinecite{Greenside81}. As ErNi$_{2}$B$_{2}$C is
strongly Type II ($\lambda /\xi \approx 5)$, we conclude that the SVP phase
is the preferred state for coexisting ferromagnetism and superconductivity.

We have measured the temperature dependence of the in-plane magnetic
penetration depth $\Delta \lambda (T)=\lambda (T)-\lambda (T_{base})$, in
single crystals of ErNi$_{2}$B$_{2}$C down to $T_{base}=0.12$~K using a
tunnel-diode based, self-inductive technique at 21~MHz \cite{Bonalde2000}
with a noise level of 2 parts in 10$^{9}$ and low drift. The magnitude of
the ac field was estimated to be less than 40~mOe. The cryostat was
surrounded by a bilayer Mumetal shield that reduced the dc field to less
than 1~mOe. \textit{The very small values of the ac and dc field in our
system ensure that our measurement is essentially a zero-field one, thereby
eliminating the possibility of trapped flux.} Details of sample growth and
characterization are described in Ref.~\onlinecite{Cho95}. The samples were
then annealed according to conditions described in Ref.~\onlinecite{Miao02}.
The sample was mounted, using a small amount of GE varnish, on a single
crystal sapphire rod. The other end of the rod was thermally connected to
the mixing chamber of an Oxford Kelvinox 25 dilution refrigerator. The
sample temperature is monitored using a calibrated RuO$_{2}$ resistor at low
temperatures (\textit{T}$_{base}$ -- 1.8~K), and a calibrated Cernox
thermometer at higher temperatures (1.3~K -- 12~K).

The deviation $\Delta \lambda (T)=\lambda (T)-\lambda (0.12$~K$)$ is
proportional to the change in resonant frequency $\Delta f(T)$ of the
oscillator, with the proportionality factor $G$ dependent on sample and coil
geometries. We determine $G$ for a pure Al single crystal by fitting the Al
data to extreme nonlocal expressions and then adjust for relative sample
dimensions \cite{Chia03}. Testing this approach on a single crystal of Pb,
we found good agreement with conventional BCS expressions. The value of
\textit{G} obtained this way has an uncertainty of $\pm $10\% because our
sample, with approximate dimensions 1.2 $\times $ 0.9 $\times $ 0.4~mm$^{3}$%
, has a rectangular, rather than square, basal area \cite{Prozorov2000}.

\begin{figure}[tbp]
\centering \includegraphics[width=8cm,clip]{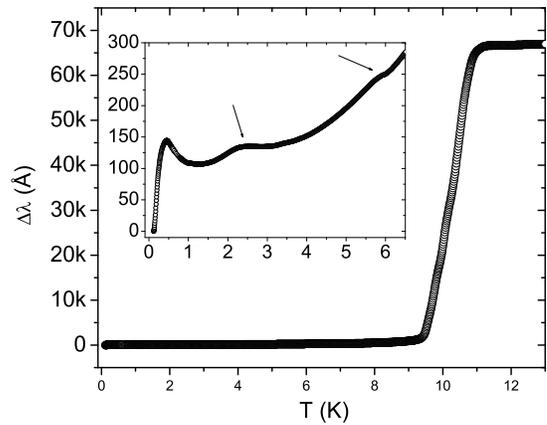}
\caption{Temperature dependence of the penetration depth $\Delta lambda$(%
\textit{T}) from 0.12~K to 13~K. Inset shows the low-temperature region. The
arrows point to the features at $T_{N}$ and $T_{WFM}$.}
\label{fig:ENBCAllT}
\end{figure}

\begin{figure}[tbp]
\centering \includegraphics[width=8cm,clip]{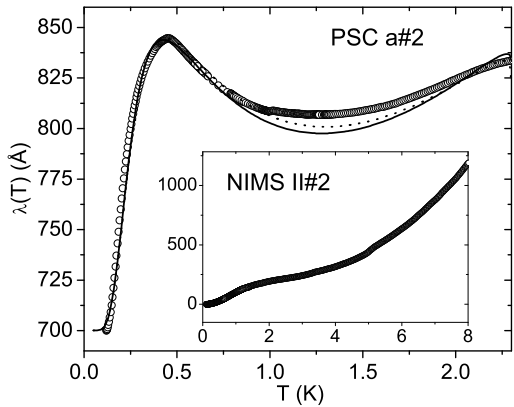}
\caption{($\bigcirc$) Temperature dependence of the penetration depth $%
\Delta \protect\lambda$(\textit{T}) in the WFM phase. We've assumed $\protect%
\lambda (T_{base}) \approx \protect\lambda $(0)=700~\AA\ from Ref.~
\onlinecite{Gammel99}. The solid line is the fit of the data to Eqn.~
\protect\ref{eqn:lambda2level} for $n$=0.21, while the dotted line is that
of $n$=0.125. The values of other parameters are mentioned in the text. The
inset shows $\Delta \protect\lambda$ for a floating-zone-grown sample
exhibiting no signals at $T_N$ or $T_{WFM}$}
\label{fig:ENBCWFM}
\end{figure}

Figure~\ref{fig:ENBCAllT} shows the temperature-dependence of the in-plane
penetration depth $\Delta \lambda (T)$. We see the following features: (1)
onset of superconductivity at $T_{c}^{\ast }=11.3$~K, (2) a slight shoulder
at $T_{N}=6.0$~K, (3) a broad peak at $T_{WFM}=2.3~$K, (4) another sharp
peak at $T_{m}=0.45~$K, and (5) an eventual downturn below $T_{m}$. The
features at $T_{N}$ and $T_{WFM}$ have not been seen in previous microwave
measurements of $\Delta \lambda (T)$ on either thin-film \cite{Andreone99a}
or single-crystal ErNi$_{2}$B$_{2}$C \cite{Jacobs95} but the former has been
observed in SANS data \cite{Gammel99}. We show in a separate publication
\cite{Chia05} that the feature at $T_{N}$ is only observed for relatively
small non-magnetic scattering rates. The large value of $T_{c}$ and the
resolvability of the features at $T_{N}$ and $T_{WFM}$ attest to the high
purity of the samples. \ We show, for comparison, data for a sample grown by
floating-zone methods in the inset to Fig. \ref{fig:ENBCWFM}. \ No clear
signal is seen at $T_{N}$, although there may be some sign of the Neel
transition near 5 K. \ In place of the up-turn in the penetration depth, the
signal levels off near $T_{WFM}$ before decreasing below 1 K. \ This
suggests that spontaneous vortices at the surface of this sample are
strongly pinned.

Figure~\ref{fig:ENBCWFM} shows the data below $T_{WFM}$. The strong upturn
is a significant deviation from the normal monotonic decrease of the
penetration depth with decreasing temperature. Because we expect the
Meissner effect to vanish ($\lambda (T)$ to diverge) in the SVP in the
absence of pinning \cite{Ng97b}, it is natural to analyze the
low-temperature data in the context of weakly pinned vortices in the
low-frequency limit. We use a two-level tunneling model proposed by Koshelev
and Vinokur (KV) \cite{Koshelev91}. This approach has been revisited by
Korshunov \cite{Korshunov01} and applied to ultrathin cuprate films by
Calame \textit{et al.} \cite{Calame01}. At relatively high frequencies,
small oscillations of the pinned lattice near equilibrium (Campbell regime)
dominate absorption. At lower frequencies, jumps of lattice regions between
different metastable states (two-level systems) come into play and determine
the absorption. Both regimes are sensitive to the pinning strength, which
depends on the value of the internal magnetic field $B_{in}$. At small
field, vortices are pinned independent of one another. When $B_{in}$ exceeds
a characteristic value $B_{p}$, pinning becomes collective \cite{Larkin79},
and the vortex lattice splits into volumes that are pinned as a whole, with
correlated regions having length $L_{c0}$ parallel to the field. We argue
below that, in the WFM phase, $H_{c1}<B_{in},B_{p}\ll H_{c2}$ and that the
frequency lies in the two-level regime.

In the two-level model, the pinning state is characterized by a large number
of neighboring metastable configurations. If we shift any volume $V$ of the
vortex lattice, then a finite probability exists that at some distance $u$
there is another state with an energy within $\Delta $ of the starting
configuration and separated from it by a barrier $U$. At finite temperatures
such regions of the lattice jump among metastable states. Under the action
of an applied ac field of frequency $\omega $, the ac screening current
exerts an ac Lorentz force on the vortex lattice which induces jumps among
nearby metastable states. The motion of the vortices, which carry magnetic
fields, then increases the ac field penetration into the sample, resulting
in an increase in the effective penetration depth $\lambda _{eff}$. The
system exhibits typical Debye-type behavior with the properties determined
by $\tau (T)\sim \tau _{0}\exp (U/T),$\ the mean time between jumps. When
two-level response dominates over Cambell behavior the penetration is
estimated \cite{Koshelev91}\ to behave as
\begin{equation}
\lambda _{eff}^{2}(T)=\lambda _{\parallel }^{2}(T)+\frac{B_{in}^{2}(T)n_{tl}%
}{16\pi T}\left\langle \frac{V^{2}u^{2}}{(1+(\omega \tau (T))^{2})\cosh
^{2}(\Delta /T)}\right\rangle ,  \label{eqn:lambda2level}
\end{equation}%
where $\langle $...$\rangle $ denotes an average over the distribution of
two level systems. Here $\lambda _{\parallel }(T)$ is the London penetration
depth in the absence of vortices; $B_{in}$, the internal magnetic field; and
$n_{tl}$, the concentration of two-level systems. In the low-field region ($%
B_{in}<B_{p}$), the vortex lines move independently, and their presence does
not change the penetration depth considerably ($\lambda _{eff}\approx
\lambda _{\parallel }$). However, in the collective pinning state ($%
B_{in}>B_{p}$), the jumping volume is not too small, and the characteristic
distance at which the nearest metastable state exists is approximately the
radius of the pinning force $u\approx \xi _{\parallel }$. As the temperature
decreases, there is insufficient thermal energy to overcome the barrier $U$.
No jumping takes place, the vortices are frozen, and hence there is no extra
penetration. One therefore recovers the London penetration depth $\lambda
_{\parallel }$ at the lowest temperatures.

The solid line shows the fit of Eq.~\ref{eqn:lambda2level} to the data below
$T_{WFM}$. In this fit, we follow KV and replace $\left\langle
...\right\rangle $ with values that characterize an effective number $%
n_{tl}^{eff}$of active two-level systems. The values of the following
quantities will be justified later: $B(T=0)\approx 1100$~G, $u\approx $ $\xi
_{\parallel }\approx 150~$\AA , $V=L_{c0}u^{2}=5.4\times 10^{-16}$~cm$^{3}$,
and $\tau _{0}=2.2\times 10^{-9}$~s. The temperature-dependence of the
internal magnetic field $B(T)$ can be obtained by fitting magnetization
values in Ref.~\onlinecite{Jensen02} to the expression
\begin{equation}
B(T)\sim \left( 1-\frac{T}{T_{WFM}}\right) ^{n}  \label{eqn:Mfit}
\end{equation}%
giving $n$=0.21. This value of $n$ is between the 2D-Ising value of 0.125
and the 3D-Ising value of 0.31, which is reasonable because in ErNi$_{2}$B$%
_{2}$C the spins lie on sheets normal to the $a$ axis and are confined to be
along or anti-along the $b$ axis, yet there is also 3D behavior in the
superconductivity. Because $U$ and $\Delta $ are strongly correlated, we
make the reasonable assumption that all metastable states are equivalent ($%
\Delta =0$) and choose the energy barrier $U=0.49$~K and pinning density $%
n_{tl}^{eff}=1.61\times 10^{11}$~cm$^{-3}$ that best fit the peak in $%
\lambda _{eff}(T)$. This value of the barrier makes $\omega \tau \approx 1$
near 1~K. Note that this value of $U$ is close to the position of the peak
at $T_{m}$ --- this is reasonable since below this temperature, the vortices
no longer have enough thermal energy to overcome the barrier to hop among
metastable states; hence, one recovers the Meissner state with $\lambda $
decreasing. We expect $\lambda _{\parallel }(T)$ to exhibit a power-law
temperature dependence at low temperatures from the combination of
gap-minima observed in non-magnetic borocarbides and the increased
pair-breaking as Er spins disorder. Consequently, we set $\lambda _{\Vert
}(T)=\lambda _{\parallel }(0)(1+bT^{2})$ with $b=0.036$~K$^{-2}$ the third
adjustable parameter in the fit. For comparison, we show the (dotted-line)
fit with $n=1/8$ (2D-Ising model) for which $U=0.49$~K, $n_{tl}^{eff}=1.57%
\times 10^{11}$~cm$^{-3}$, and $b=0.033$~K$^{-2}$. Both fits reproduced the
qualitative features of the data, though the latter curve fits the data
slightly better.

To justify our application of the two-level model to our data, we evaluate
various physical parameters in the model using standard expressions for the
vortex state \cite{Blatter94}. We start with the measured quantities:
zero-temperature in-plane penetration depth $\lambda _{\parallel }(0)\approx
700$~\AA\ and coherence length $\xi _{\parallel }\approx 150$~\AA\ (and $%
\kappa =\lambda _{\parallel }(0)/\xi _{\parallel }=4.7)$\ from Ref.~%
\onlinecite{Gammel99}, ferromagnetic moment $M\approx 0.62\mu _{B}$/Er $%
\approx $ 100~Oe well below $T_{WFM}$ from Ref.~\onlinecite{Gammel00}, and
the normal-state resistivity $\rho _{n}(T_{c}^{\ast })=5.8$~$\mu \Omega $ cm
that we measured for our sample. The value of $M$ corresponds to an internal
field $B\sim 4\pi M=1100$~G, which is greater than $H_{c1}\sim 500$~G,
putting us in the mixed state.

In Table I, we give the expressions and values for the quantities that lead
to the flux coherence length $L_{co}$ ($19.5$ nm), the collective pinning
field $B_{p}$ (20 Oe), the frequency $\omega _{cr}$ above which Campbell
response is expected (27~MHz), and the jump-time prefactor $\tau _{0}$
(2.2~ns). Since we operate at 21~MHz, this puts us in the two-level regime.
Note that $B_{p}$ is less than H$_{c1},$ suggesting the the mixed state of
ErNi$_{2}$B$_{2}$C is always in the collective pinning regime. Based on
these values, we estimate the maximum density of two-level volumes to be $%
\sim 2.4\times 10^{13}$cm$^{-3}$ at the lowest temperatures, indicating that
approximately 1\% are active in our frequency window.

\begin{widetext}
\begin{table}[tbp]
\centering%
\begin{tabular}{llll}
\hline
{\small Quantity} & {\small Expression} & {\small Value} & {\small Notes} \\
\hline\hline
{\small Depairing current, }$j_{s}$ & ${\small c\Phi }_{0}{\small /(12}\sqrt{%
3}{\small \pi }^{2}{\small \xi }_{\parallel }{\small \lambda }_{\parallel
}^{2}{\small )}$ & $1.5\times 10^{8}${\small ~A/cm}$^{2}$ &  \\
{\small Viscous drag coefficient, }$\eta $ & ${\small \approx \Phi }_{0}^{2}%
{\small /(2\pi \xi }_{\parallel }^{2}{\small \rho }_{n}{\small c}^{2}{\small %
)}$ & $5.6\times 10^{-7}${\small ~erg~s~cm}$^{-3}$ & {\small %
Bardeen-Stephens model} \\
{\small Bean-model critical current, }$j_{c}$ & ${\small 4cM}_{h}{\small /L}$
& $1.9\times 10^{4}${\small ~A/cm}$^{2}$ & {\small Ref.~\cite{Gammel00}, }$%
L=1${\small \ mm,} \\
&  &  & $M_{h}${\small \ from hysteresis loop} \\
{\small Flux coherence length, }$L_{c0}$ & $x^{2}=\frac{j_{s}}{j_{c}}\ln x;$%
{\small \ }$x=\frac{\lambda _{\perp }L_{c0}}{\lambda _{\parallel }\xi
_{\parallel }}$ & {\small 19.5 nm} & {\small Ref.~\cite{Cho95},\cite%
{Koshelev91}, }$\lambda _{\perp }/\lambda _{\parallel }\approx 1.3$ \\
{\small Jump time prefactor, }$\tau _{0}$ & ${\small \eta \xi }_{\parallel }%
{\small c/(\Phi }_{0}{\small j}_{c}{\small )}$ & $2.2\times 10^{-9}${\small %
~s} & {\small Ref.~\cite{Koshelev91}} \\
{\small Collective pinning field, }$B_{p}$ & ${\small \Phi }_{0}{\small (}%
\ln {\small x)}^{4/3}{\small /L}_{c0}^{2}$ & {\small 20 Oe} & {\small Ref.~%
\cite{Koshelev91}} \\
{\small Campbell crossover, }$\omega _{cr}(B,T)$ & ${\small \approx T\Phi }%
_{0}{\small /(\eta BV\xi }^{2}{\small )}$ & $27${\small ~MHz} & {\small Ref.~%
\cite{Koshelev91}, }$B${\small =500~Oe; }$T${\small =2~K} \\ \hline
\end{tabular}%
\caption{Vortex parameters for the two-level hopping regime}
\end{table}
\end{widetext}

KV \cite{Koshelev91} also found the Campbell penetration depth to be $\sim $$%
B^{2}$, which is a monotonically increasing function with decreasing
temperature, i.e. there is no peak at low temperatures. This is in agreement
with our not being in the Campbell regime. We also measured $\lambda (T)$
with the ac field along the basal plane, finding features qualitatively
similar to the present data, including the strength and position of the
features at $T_{N}$, $T_{WFM}$ and $T_{m}$.

In conclusion, penetration depth data of single-crystal ErNi$_{2}$B$_{2}$C
down to $\sim $0.1~K provide strong evidence for the existence of a
spontaneous vortex phase below $T_{WFM}$. The high quality of our sample
enables us to see features at $T_{N}$ and $T_{WFM}$ that have not been
observed in previous studies of the penetration depth \cite%
{Andreone99a,Jacobs95}. Other samples, such as that shown in the inset to
Fig. \ref{fig:ENBCWFM} show no clear signal at either $T_{N}$ or $T_{WFM}$,
nor the upturn in penetration depth that we attribute to weakly pinned
spontaneous vortices. As pointed out by Radzhiovsky \cite{radzihovsky01},
the SVP lattice is much softer than a conventional lattice, and therefore
especially sensitive to quenched disorder. It may well be that the
spontaneous vortices may be glass-like rather than forming a lattice, and
that ferromagnetic closure domains form at the surface. While these aspects
may make spontaneous vortices difficult to detect by neutron scattering or
surface magnetization, vortices will still have strong effects on the
electrodynamics, as observed here.

This material is based upon work supported by the U.S. Department of Energy,
Division of Materials Sciences under Award No. DEFG02-91ER45439, through the
Frederick Seitz Materials Research Laboratory at the University of Illinois
at Urbana-Champaign. Research for this publication was carried out in the
Center for Microanalysis of Materials, University of Illinois at
Urbana-Champaign. This work in Korea was supported by the Ministry of
Science and Technology of Korea through the Creative Research Initiative
Program.

\bibliographystyle{prsty}
\bibliography{RNBC,CeCoIn5,Vortex}

\end{document}